\documentclass[10pt]{iopart}

\usepackage{iopams}  

\usepackage{graphicx}
\begin{document}

\title[]{Numerical analysis of surface and edge states in slabs, 
stripes, rods and surface steps of topological insulators}

\author{N.I.~Fedotov, S.V.~Zaitsev-Zotov}

\address{Kotel'nikov IRE RAS, Mokhovaya 11, bld.7, 125009 Moscow, Russia}
\ead{nfedotov89@mail.ru}
\vspace{10pt}
\begin{indented}
\item[]\today
\end{indented}

\begin{abstract}
By  numerically solving the effective continuous model of a topological insulator with parameters corresponding to the band structure of the topological insulator Bi$_2$Se$_3$, we analyze 
possible appearance of one-dimensional states in various geometries.  
Massless Dirac fermions are found at the edges of thin ribbons with surface oriented not only along the van der Waals gap but also  in the perpendicular direction. Thick rods and slabs with surface steps host massive modes localized on surface faces. We argue that the modes are massive and their origin  is due to the difference  in the Dirac point energy of adjacent faces. The absence of one-dimensional  states near edges of a large rectangular rod and surface steps is demonstrated.
\end{abstract}

\vspace{2pc}

\maketitle

\ioptwocol

\section{Introduction}
As  is now well known, dielectrics are characterized not only by the energy gap, but also by a topological invariant $\mathbb{Z}_2$, the nonzero value of which leads to the appearance of surface states protected by symmetry \cite{Z2}. Materials with non-zero $\mathbb{Z}_2$ are called topological insulators (TI) and have attracted a great interest in the last years \cite{reviews}. The surface states have Dirac-like energy spectrum and their spin direction depends on their momentum direction (spin-momentum locking). Spin-momentum locking inspires a hope for the emergence of almost non-dissipative electronic transport in TI.
In this regard, the most interesting and promising systems are the ones with 1D topologically-protected states. 
The number of such systems is rapidly increasing. They include semiconductor heterostructures (CdHgTe \cite{Molenkamp} and InAs/GaSb \cite{InAsGaSb} systems), various realizations of graphene-like structures 
 \cite{analogs}, steps on the surface of crystalline TI \cite{crystalTI}, thin layers of some Weyl semimetals \cite{Weyl1,Weyl2,Weyl3} and magnetically doped TI \cite{magneticTI}.

Bi$_2$Se$_3$ is a prototypical TI with a large bulk band gap and a single Dirac cone with negligible warping and the Dirac point in the bulk gap \cite{Zhang2009}.
Bismuth and antimony chalcogenide based nanostructures are popular objects of experimental \cite{Peng, Xiu2011, Tian2013, Hong2014, Cha2012,  Ning2013, Li2012, Li2014, Baessler2015} and theoretical \cite{Bardarson2010, Zhang2010, Egger2010, Brey, Tightbinding} research due to their large surface to volume ratio and therefore a smaller contribution of the bulk to transport properties. Nanostructure of a wide range of shapes (nanoplatelets, nanowires, straight and zig-zag nanoribbons) can be obtained by a variety of techniques \cite{Kong2010, Knebl2014, LiH2012, Zou, Baessler2015}. Phenomena experimentally observed in these structures   include Aharonov-Bohm \cite{Peng, Xiu2011, Tian2013} and  Altshuler−Aronov−Spivak oscillations \cite{Tian2013}, weak antilocalization \cite{Cha2012,  Ning2013} and universal conductance fluctuations \cite{Li2012, Li2014}.

A 2D topological insulator phase and 1D symmetry-protected edge states were predicted for thin layers of Bi$_2$Te$_3$ with odd number of quintuple layers \cite{oscillations1}, in thin layers of Bi$_2$Se$_3$ \cite{oscillations,Shan}.
There is a number of predictions and speculations on possible realizations of various types of 1D states near edges connecting faces of a TI. Existence of 1D edge states along a junction between two topological insulator surfaces was predicted on the basis of 2D Dirac equations for the surface states \cite{Deb}. They disappear in a more accurate consideration \cite{Brey}  and are recovered upon introduction of a delta function potential on the edge \cite{rib1}.
The existence of 1D states localized at surface steps may be expected in systems where the Fermi velocity of surface states assumes different values on different faces of the surface of the TI \cite{Moon}, in analogy with evanescent waves in optical wave-guiding. 
1D edge states were predicted for steps on the surface of Bi$_2$Se$_3$ in DFT calculations \cite{Narayan2014} and on the surface of Bi$_2$Te$_3$ in tight-binding simulations \cite{Kobayashi}.

Experimental verification of such predictions provides controversial results. On the one hand, $\approx 20$\% growth of LDOS near a surface step edge was observed  in Bi$_2$Te$_3$ in scanning-tunneling spectroscopy (STS) experiments \cite{Alpichshev}. The effect was initially explained as appearance of bound states. On the other hand, even much bigger increase of LDOS  in the Dirac point in Bi$_2$Se$_3$ is observed in STS experiments \cite{FedotovPRB}. This increase is accompanied by shift of the chemical potential level by $\approx 0.2$ eV. It is shown that such a shift increases the normalized tunneling conductance \cite{FedotovJETPL} and gives illusion of the edge states. A more careful analysis of the experimental data accompanied by numerical simulation reveals however the emergence of  bound 1D edge states in the potential well formed due to chemical potential shift near surface steps \cite{FZZ2018}.

We focus our attention here on slabs, 
stripes, rods and surface steps of Bi$_2$Se$_3$. It is chosen as a model topological insulator for its simple energy structure. The results obtained retain general validity for other topological insulators, although they may differ in details and be complicated by the presence of warping, magnetic field, a potential barrier, {\it etc}. 

The properties of nanostructures and surface steps can ba analyzed using DFT \cite{Narayan2014}, tight-binding computations \cite{Kobayashi, Tightbinding,Xu2018} or effective Hamiltonians \cite{Brey}. The latter technique requires less parameters and can be utilized for the investigation of larger structures. While not as accurate, it allows us to capture the essential physics without concentrating on the structural details. In contrast with other studies, which use a 2D effective Hamiltonian and model the surface step or the nanorod edge as a $\delta$-function barrier \cite{Biswas2011, Liu2012, Zheng2012, An2012}, we choose a 3D one to account for the 3D nature of the problem.  

The aim of the present work is to clarify the following questions: whether 1D states appear in TI near surface edges in various geometries;  whether side surfaces or faces of surface steps host bound surface states. The problems are analyzed by numerically solving the effective continuous model proposed by Zhang {\em et al.} \cite{Zhang2009} with parameters corresponding to the band structure and surface states of bulk Bi$_2$Se$_3$. We demonstrate here that in the framework of the chosen model the 2DTI phase and massless Dirac fermions at the edges are present  in thin ribbons with surface oriented not only along the van der Waals gap, but also in the perpendicular direction.  Absence of one-dimensional massless Dirac states near edges of a large rectangular rod and surface steps is demonstrated. Instead we  find that thick rods and slabs with surface steps host massive modes localized on surface faces due to a difference in Dirac point position for adjacent faces. The picture in the rods with small steps is complicated by the finite penetration length of the surface steps.

\section{Model and calculation methods}
In the framework of the effective continuous model proposed by Zhang {\em et al.} \cite{Zhang2009} the states of a TI near the $\Gamma$-point can be described by a Hamiltonian.

\begin{equation}
H= E_0(\bi{k})+
 \left(\begin{array}{rrrr}
  M(\bi{k}) & A_1k_z & 0 & A_2k_-\\
  A_1k_z & -M(\bi{k}) & A_2k_- & 0\\
  0 & A_2k_+ & M(\bi{k}) & -A_1k_z\\
  A_2k_+ & 0 & -A_1k_z & -M(\bi{k})\\
 \end{array}\right),
 \label{eq:ham}
\end{equation}
where
$k_{\pm}=k_x\pm ik_y$
$E_0(\bi{k})=C+D_1k_z^2+D_2(k_x^2+k_y^2),$ $M(\bi{k})=M-B_1k_z^2-B_2(k_x^2+k_y^2)$ and $A_1$, $A_2$, $B_1$, $B_2$, $C$, $D_1$, $D_2$ are numerical parameters. 
We  use the following set of parameter values corresponding to Bi$_2$Se$_3$ energy structure: 
$M=0.28$ eV,
$C=-0.0068$ eV,
$A_1=0.22$ eV \AA{},
$A_2=0.41$ eV \AA{},
$B_1=0.1$ eV \AA{}$^2$,
$B_2=0.566$ eV \AA{}$^2$,
$D_1=0.013$ eV \AA{}$^2$,
$D_2=0.196$ eV \AA{}$^2$ \cite{Zhang2009}. 
Here  $z$ is the direction  transverse to the cleavage surface of Bi$_2$Se$_3$ and the  $x$--$y$ plane is parallel to it.

We will study here the energy structure of slabs, rods and ribbons, and steps with various orientations.  
In the case of rods, ribbons and steps running along the $y$  axis
we are considering systems translationally invariant along the $y$  axis, so $k_y$  is conserved and two remaining wave vector components are replaced by their operators $k_x\rightarrow-i\partial_x$ and $k_z\rightarrow-i\partial_z$. For each $k_y$   we treat the corresponding resulting Dirac equation as a two-dimensional equation for a continuous wave function $\psi^{(\alpha)}(x,z)$, where $\alpha=1,...,4$ is the index of the wave function component, and solve it by a standard finite difference method. Specifically, we introduce a rectangular grid $x_n=nh_x$, $z_m=mh_z$, where $n=0,...,N-1$, $m=0,...,M-1$, typically $h_x=0.4{\rm -}0.5$ nm, $h_z=0.2{\rm -}0.4$ nm and $N \times M=1000{\rm -}3000$. Note that this grid is not connected with the actual crystalline lattice of the  material under consideration. Then we discretize the wave function $\Psi^{(\alpha)}_{n,m}=\psi^{(\alpha)}(x_n,z_m)$. The differential operators are replaced by central finite differences 
\begin{eqnarray*}
\partial_x\psi^{(\alpha)}(x_n,z_m)\rightarrow(\Psi^{(\alpha)}_{n+1,m}-\Psi^{(\alpha)}_{n-1,m})/2h_x,\\
\partial_z\psi^{(\alpha)}(x_n,z_m)\rightarrow(\Psi^{(\alpha)}_{n,m+1}-\Psi^{(\alpha)}_{n+1,m-1})/2h_z,\\
\partial_{xx}\psi^{(\alpha)}(x_n,z_m)\rightarrow(\Psi^{(\alpha)}_{n+1,m}-2\Psi^{(\alpha)}_{n,m}+\Psi^{(\alpha)}_{n-1,m})/h_x^2,\\
\partial_{zz}\psi^{(\alpha)}(x_n,z_m)\rightarrow(\Psi^{(\alpha)}_{n,m+1}-2\Psi^{(\alpha)}_{n,m}+\Psi^{(\alpha)}_{n,m-1})/h_z^2,
\end{eqnarray*}
We model the surface with zero boundary conditions.
For the rods and ribbons that means putting $\Psi^{(\alpha)}_{-1,m}\equiv\Psi^{(\alpha)}_{N,m}\equiv\Psi^{(\alpha)}_{n,-1}\equiv\Psi^{(\alpha)}_{n,M}\equiv0$ in the above expressions.
For the surface steps of height $L_s=Sh_z$ the boundary conditions look as follows
\begin{eqnarray*}
\Psi^{(\alpha)}_{-1,m}\equiv0, m=0,...,S-1\\
\Psi^{(\alpha)}_{-1,m}\equiv\Psi^{(\alpha)}_{N-1,m-S}, m=S,...,M-1\\
\Psi^{(\alpha)}_{N,m}\equiv\Psi^{(\alpha)}_{0,m+S}, m=0...M-S-1\\
\Psi^{(\alpha)}_{N,m}\equiv0, m=M-S,...,M-1\\
\Psi^{(\alpha)}_{n,-1}\equiv\Psi^{(\alpha)}_{n,M}\equiv0
\end{eqnarray*}
This reduces the problem to a system of linear algebraic equations for $4NM$ variables $\Psi^{(\alpha)}_{n,m}$.
The LDOS is defined as $\rho(\bi{r},E)=\sum_i|\psi_i(\bi{r})|^2\delta(E-E_i)$, where $\psi_i(\bi{r})$ and $E_i$ are the wave function and the energy of the $i$-th state. Partial LDOS is obtained by summation not over all values of $i$ but a subset thereof. 
The treatment of  rods, ribbons and steps running along the $z$  axis is analogous. 
In the case of an infinite slab two momentum projections are good quantum numbers and we solve only a one-dimensional equation numerically.

\section{Results}
\subsection{Infinite slab}
Analysis of the properties of a slab allows us to verify the method and the approximation used. 

\begin{figure}
\begin{center}
\includegraphics[width=4.2cm]{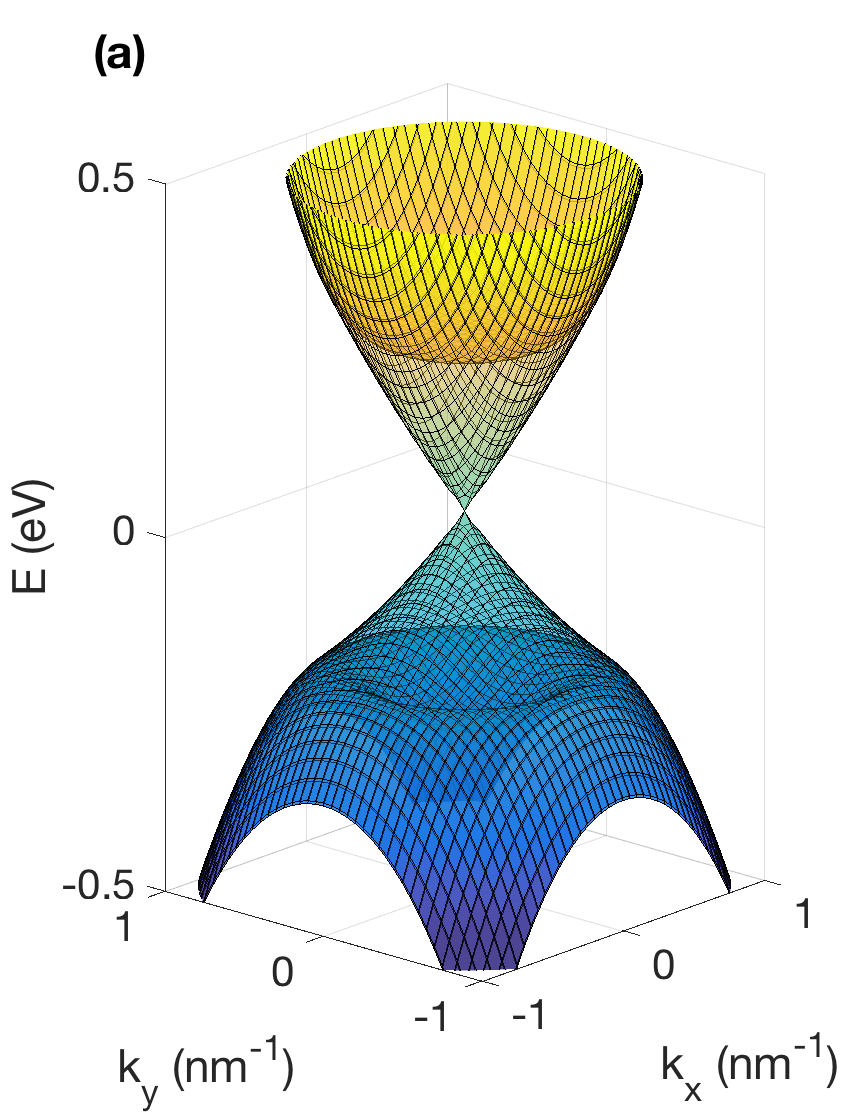}
\includegraphics[width=4.2cm]{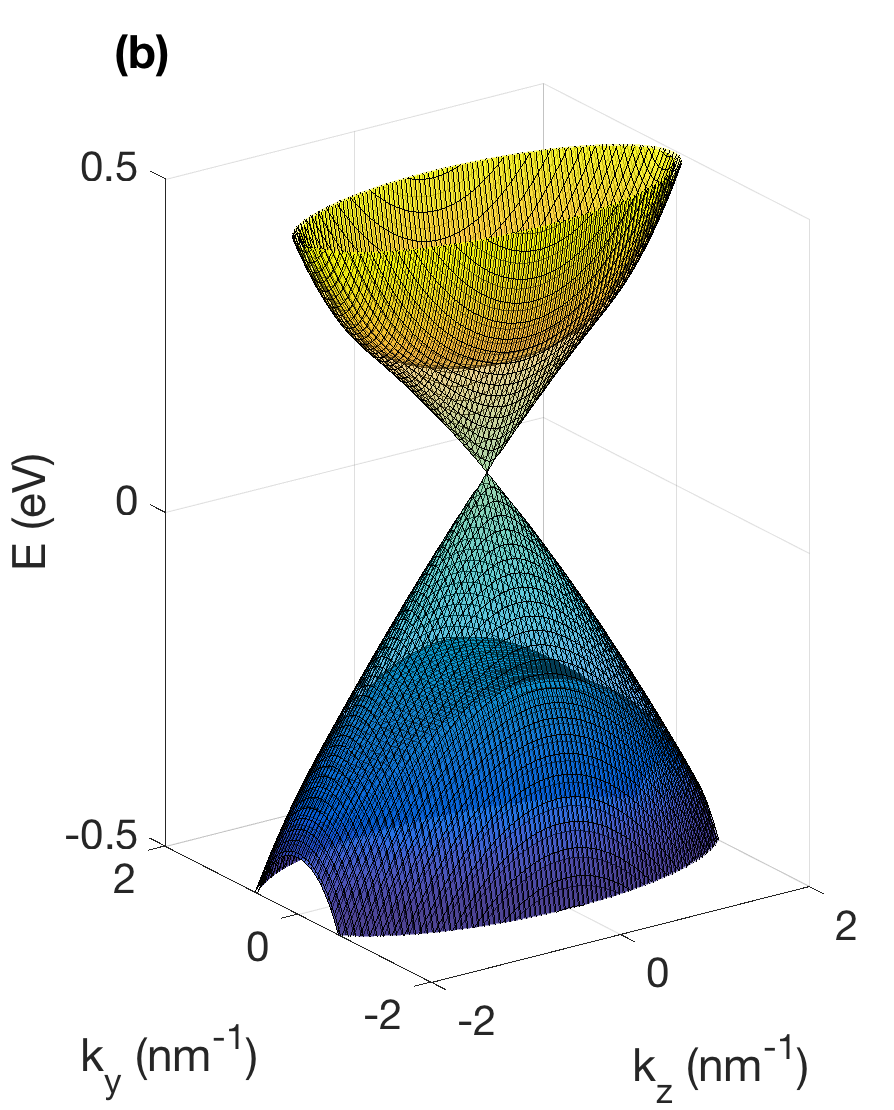}
\caption{Dispersion laws for a 25~nm thick slabs with non-overlapping surface states. (a) surface orientation along the $x$--$y$ plane; (b) along $y$--$z$ plane.
}
\label{widefig3}
\end{center}
\end{figure}

Fig.~\ref{widefig3} shows the dispersion curves for the surface and bulk states obtained by solving Equation~\ref{eq:ham} for different surface orientations for a 25 nm thick slab. Such a thickness is sufficient to reduce the effect of surface state hybridization to a negligible level. 
We have here three different energy regions: valence band states region $E\lesssim - 0.19$ eV, surface states region $-0.19\lesssim E\lesssim 0.28$ eV, and the conduction band region $E\gtrsim 0.28$ eV.  
The Fermi velocity   for $x$--$y$ plane $v_F = A_2 \sqrt{1-(D_1/B_1)^2}\approx 0.4$ eV$\cdot$nm \cite{Shan} 
 is close to the value observed experimentally \cite{ARPES} whereas the bulk energy gap is slightly above the experimental one (0.3~eV~\cite{ARPES,STM}). Another difference is the position of the Dirac point inside the bulk energy gap: it is located approximately in the middle of the energy gap for the $x$-$y$ face, whereas  both ARPES and STS measurements give approximately  0.1~eV above the valence band. 
Anisotropy of the energy structure  manifests itself in the elliptic shape of the constant-energy surfaces (Fig.~\ref{widefig3}(b)). The model also reproduces different Dirac point positions on different faces, in agreement with {\em ab initio} calculations \cite{Moon}.

\begin{figure}
\begin{center}
\includegraphics[width=7cm]{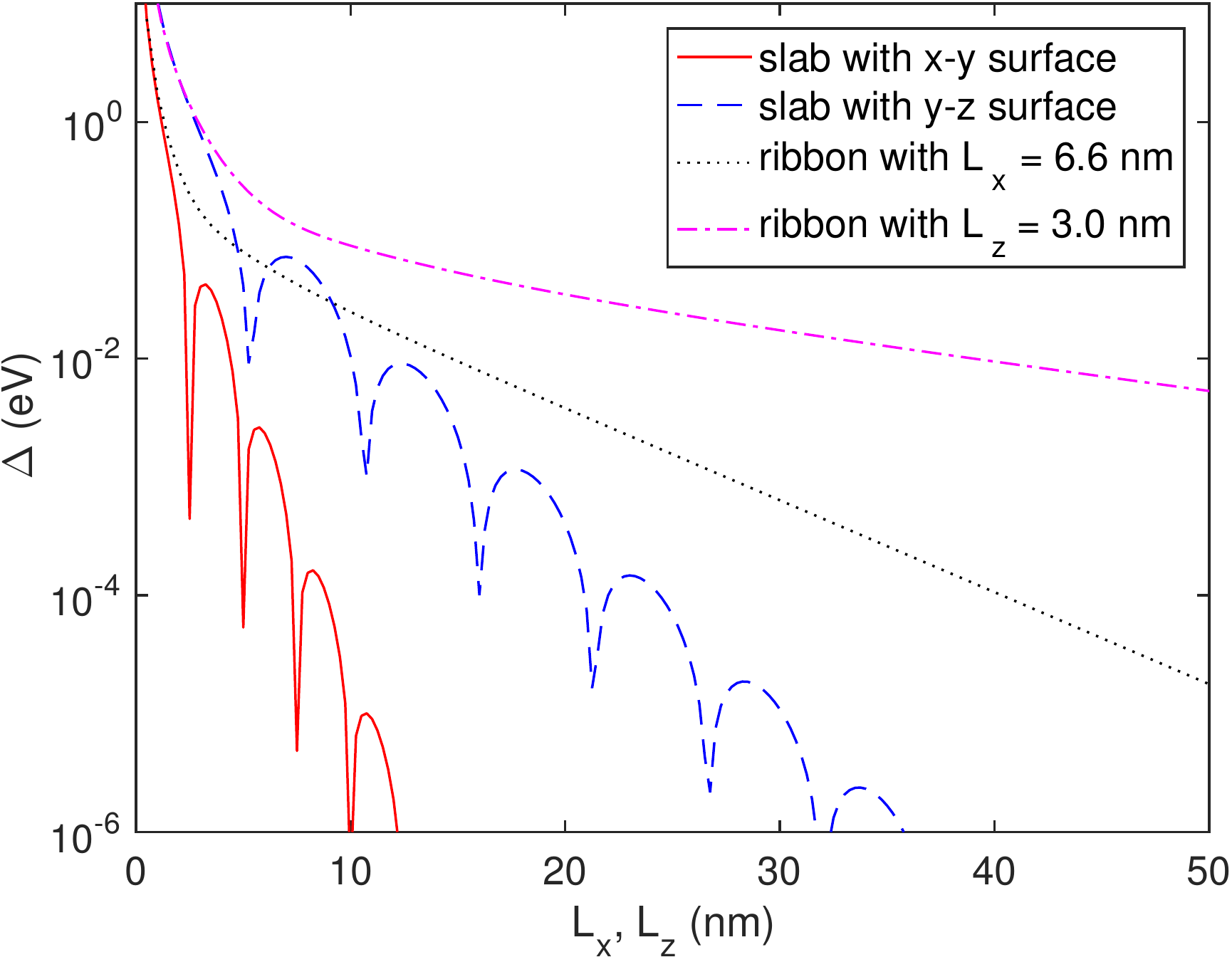}
\caption{Energy gap as a function of the slab thicknesses (blue and red curves) and ribbon width for critical thickness ribbons.}
\label{decay}
\end{center}
\end{figure}

Decrease in the slab thickness leads to hybridization of the surface states. As a result, an energy gap is opened for the surface states. The bulk energy gap also increases due to the quantum size effect. Fig.~\ref{decay} shows the gap value as a function of the slab thickness for the slab surface orientations along the $x$-$y$ (a) and $x$-$z$ (b) planes. Slower gap decrease with slab thickness along the $x$ (and $y$) direction corresponds to bigger decay length of the surface states in this direction. Periodical modulation of the gap value results from nonzero imaginary parts in the exponents $\lambda_1$ and $\lambda_2$ defining the decay of the surface states wave functions $\Psi(z)\propto e^{\lambda_1/z}+ e^{\lambda_2/z}$ in the effective continuous model \cite{Shan}. Such oscillations correspond to periodical band inversion \cite{oscillations} and are responsible for the development of a 2D topological insulator phase in  certain critical regions of slab thickness (see below).

\subsection{Rectangular rod}
\begin{figure}
\begin{center}
\includegraphics[width=7cm]{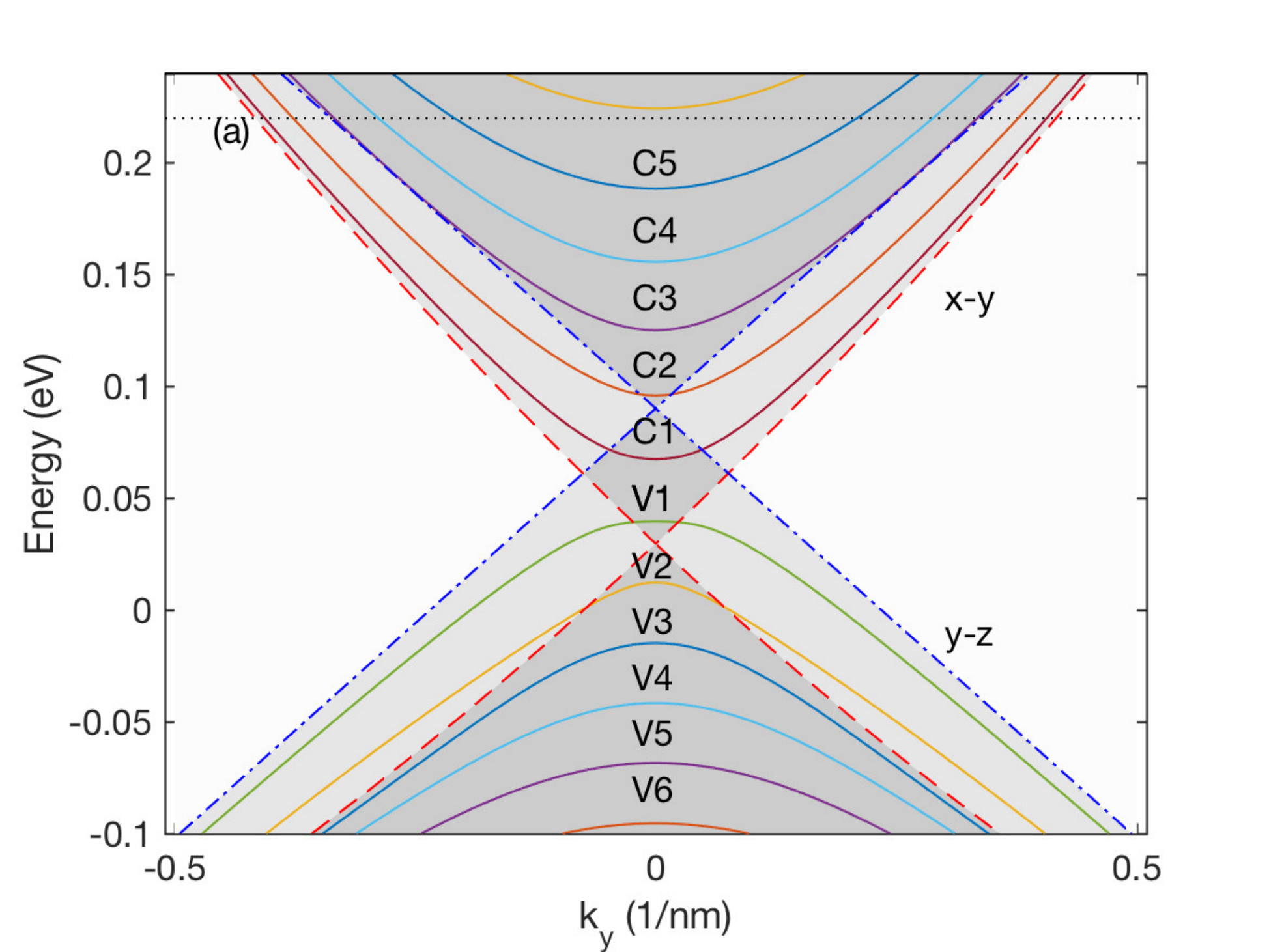}
\includegraphics[width=3.5cm]{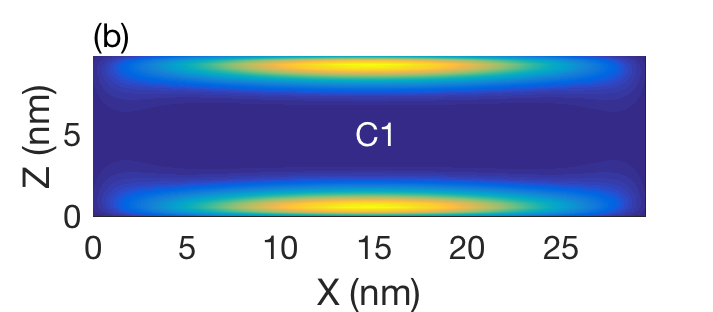}
\includegraphics[width=3.5cm]{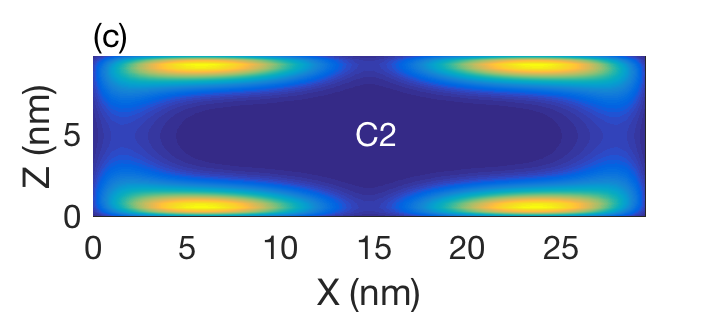}
\includegraphics[width=3.5cm]{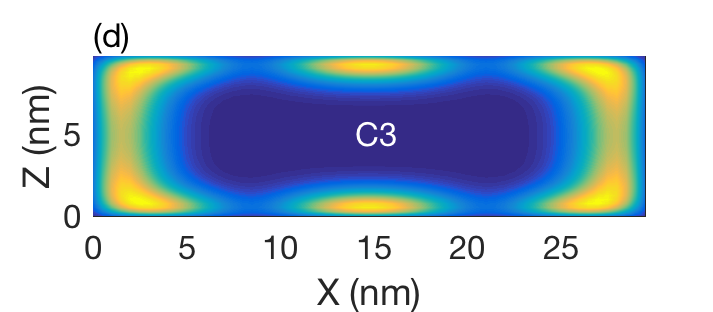}
\includegraphics[width=3.5cm]{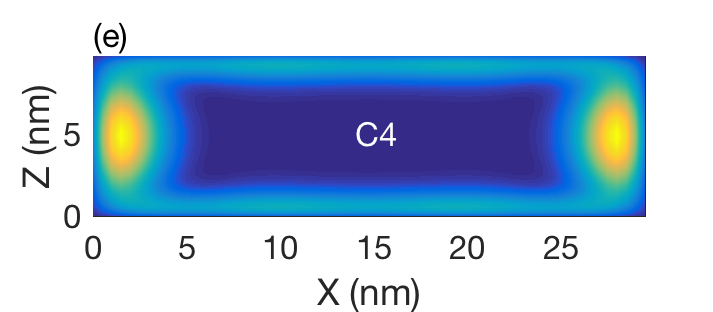}
\includegraphics[width=3.5cm]{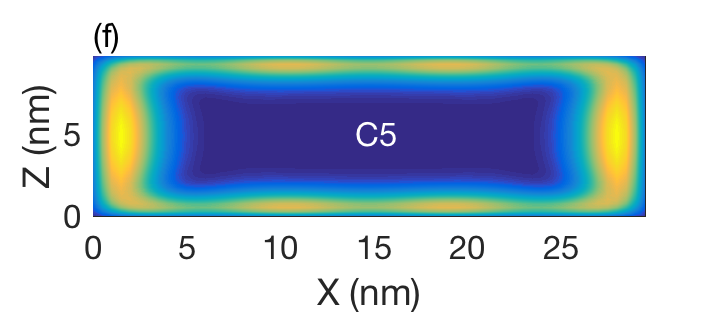}

\caption{(a)  Dispersion law for a $10\times 30$ nm$^2$ rod in the bulk energy gap region. Each mode is doubly degenerate. (b)--(f) Spatial distribution of partial local density of states (LDOS) across the rod for all 5 modes at $E=0.22$~eV (shown by dotted line). Dashed and dash-dotted lines correspond to $E(k_x=0,k_y)$ and $E(k_y,k_z=0)$ for $x$--$y$ and $y$--$z$ faces respectively.}
\label{rodfig3}
\end{center}
\end{figure}

Analysis of the states of a rectangular rod allows us to clarify the question of possible appearance of  edge states in thin layers of topological insulators, as well as in the corners of a thick and wide one. Let us consider a rod directed along the $y$ axis. $k_x$ is now not a good quantum number any more, but $k_y$ is. Fig.~\ref{rodfig3}(a) shows the dispersion curves for electron states in a $10\times 30$ nm$^2$ rod.  
Here both the rod's thickness and width are big enough to exclude overlapping of the surface states of opposite faces. 
So the bulk energy gap is approximately the same as in the case of a thick slab, but the surface states are quantized and a small energy gap develops. 

\begin{figure}
\begin{center}
\includegraphics[width=7cm]{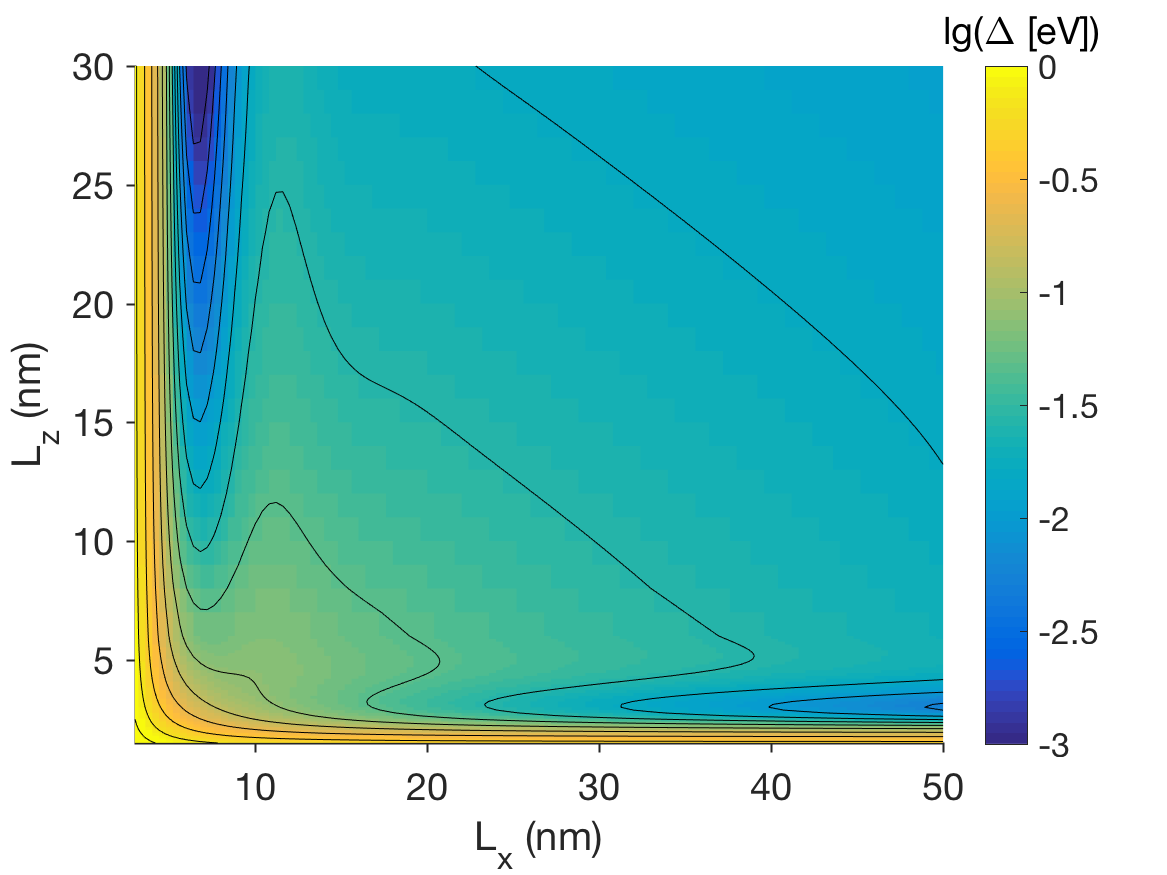}
\caption{Energy gap (logarithmic scale) in the $\Gamma$ point as a function of the rod sizes.}
\label{gapxy}
\end{center}
\end{figure}

The quantization is determined by the effective perimeter of the rod cross-section and obeys the following equations

\begin{equation}
\left\{
\eqalign{
L_x k_x+ L_z k_z=\pi \left(N+\frac{1}{2}\right)\\
k_xv_{Fx}+E_{Dxy}=k_zv_{Fz}+E_{Dyz},}\right.
\end{equation}
accounting for face- and direction-dependent $v_F$, Berry phase and dependence of the Dirac point position on face orientation. For the sake of simplicity we neglect here the weak energy dependence of $v_F$. 
The difference between the energy quantization levels of the surface states with small wavevectors is then 
\begin{equation}
\Delta E=\frac{ \pi v_{Fx}v_{Fz}}{v_{Fz} L_x +v_{Fx}L_z}.
\label{eq:delta}
\end{equation}
Alternatively, it can be rewritten as 
\begin{equation}
\Delta E=\frac{2 \pi v_{Fx}}{P_{eff}},
\label{eq:perimeter}
\end{equation}
where  $P_{eff}=2L_x+2\frac{v_{Fx}}{v_{Fz}}L_z$  is the effective perimeter of a rod. This is an approximate formula not taking into account the depth distribution of the surface states.

Fig.~\ref{gapxy} maps the value of the energy gap in the $\Gamma$ point {\em vs.} $L_x$ and $L_z$. The map consists of a relatively flat plateau at ($L_x\gtrsim15$ nm, $L_z \gtrsim 5$nm) and two deep gorges along the $x$ and $z$ axes resulted from the oscillating character of the energy gap. The plateau has a slope shown by the level lines. They correspond to constant values of the effective perimeter, $L_x+1.8 L_z={\rm const}$, in agreement with equation \ref{eq:perimeter}.  Calculations for the rod running in the $z$ direction give analogous results.

As  noted above, the Dirac point position depends on face orientation. In the model discussed here this splitting is $ \approx 0.06$~eV (Fig.~\ref{rodfig3}). 
As a result, the motion of the Dirac electrons from face to face occurs through a set of rectangular potential wells. Therefore, electron states 
can be bound to certain faces \cite{FZZ2018}. 

This behavior is illustrated by Fig.~\ref{rodfig3}(b) showing partial LDOS at $E=0.22$~eV. There are five different modes at this particular energy. 
Parts of the modes C1, C2  in Fig.~\ref{rodfig3}(a) which are not inside the Dirac cone of the $y$-$z$ face (light gray area in Fig.~\ref{rodfig3}(a)) are confined to the $x$--$y$ face.
Other modes (C3,C4, ...) belong to both cones (dark gray area in Fig.~\ref{rodfig3}(a)) and are distributed along the entire rod perimeter exhibiting a usual resonance structure. No massless Dirac mode is present in this geometry. Similar behavior is observed for V1,V2,... modes, but now V1 and V2 spread over the $y$--$z$ face.

\begin{figure}
\begin{center}
\includegraphics[width=7cm]{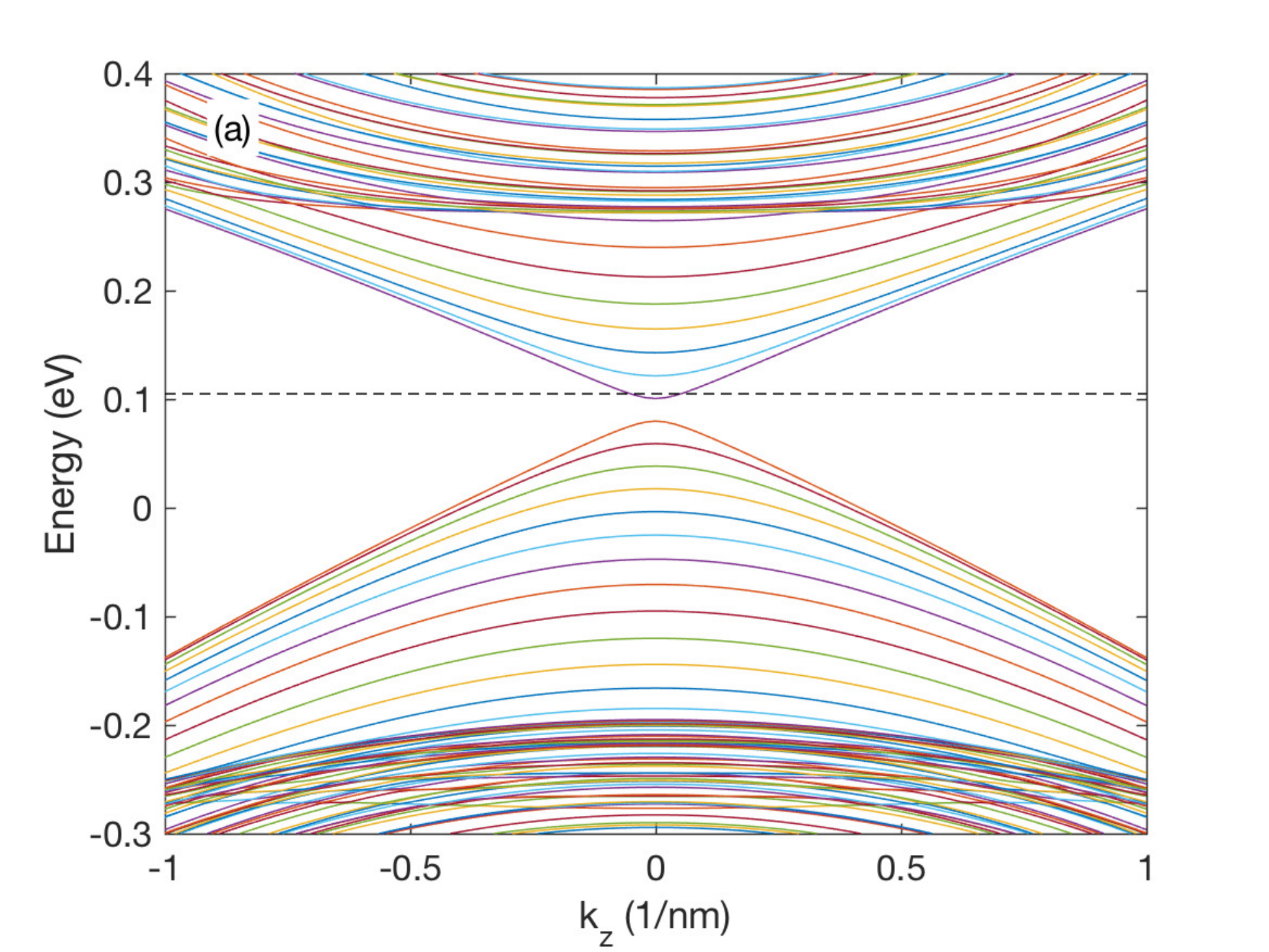}
\includegraphics[width=4cm]{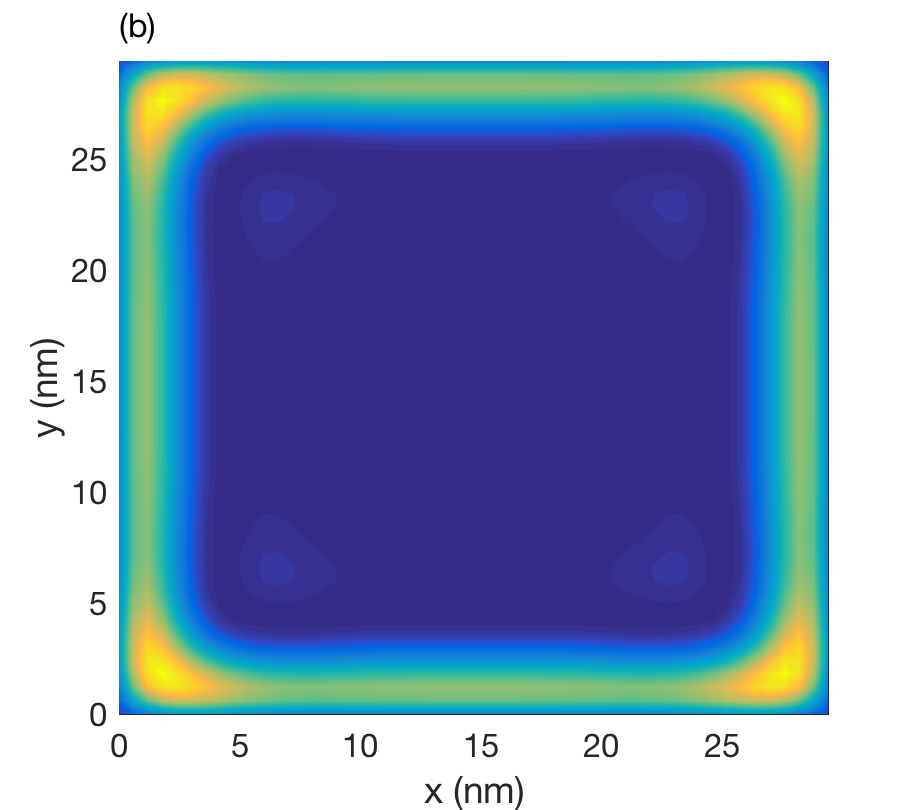}
\caption{(a) Energy structure of a $30\times 30$ nm$^2$ rod. Each mode is doubly degenerate. (b) Spatial distribution of LDOS  across the rod  at $E=0.105$~eV (shown by dotted line)}
\label{rodxy}
\end{center}
\end{figure}

In the case of a rod oriented in the $z$ direction, the Dirac point position is the same for all its faces. As a consequence, no face-specific surface states appear in this case (Fig.~\ref{rodxy}). Fig.~\ref{rodxy}(b) shows an increase in LDOS near the edges of the rod, but it does not correspond to any bound edge states, as is evident from Fig.~\ref{rodxy}(a). Indeed, all the modes in  Fig.~\ref{rodxy}(a) are inside the Dirac cone for the $x$-$z$ (and $y$-$z$) face and are therefore distributed over all of the surface of the rod. The gap in the surface states is analogous to the one for the rod, oriented in the $y$ direction.

\subsection{2D topological insulator state in a thin ribbon}
Another question of interest to us is a possibility of appearance of 1D edge states in a thin rod (ribbon) of a topological insulator. Oscillating character of the slab energy gap (see Fig.~\ref{decay}) indicates periodical  energy gap inversion leading to the development of such states. 
In Fig.~\ref{gapxy}  we see two deep gorges along $x$ and $z$ axes, the deepest one for $L_x\approx 6.6$~nm, and the next one for $L_z\approx 3.0$~nm. The gorges relate to the first regions of the gap inversion along respective directions. Gorges corresponding to other inversion regions of thickness are 
negligible due to much larger decay lengths of the edge states (see below) and have therefore no practical interest.

\begin{figure}
\begin{center}
\includegraphics[width=7cm]{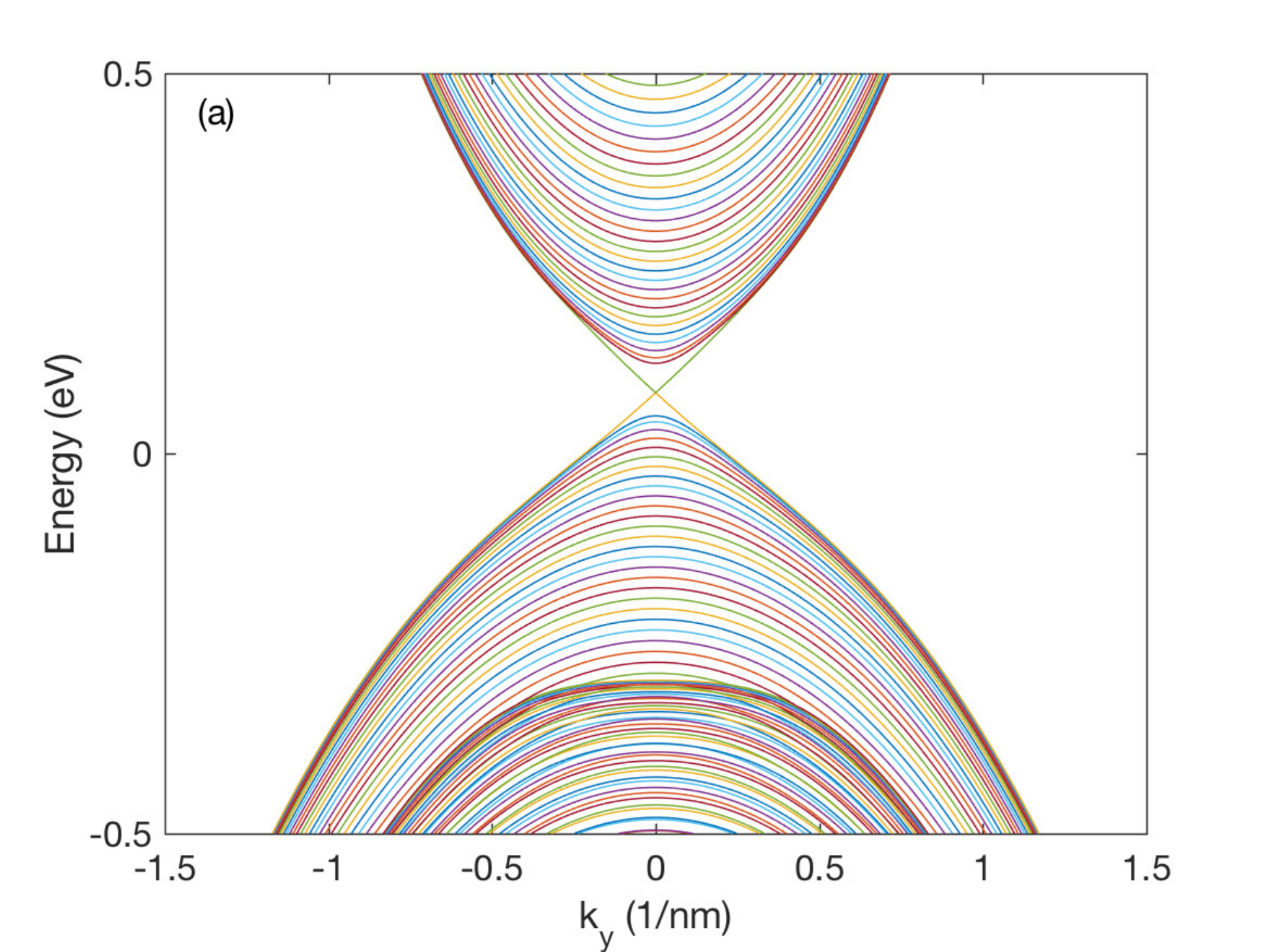}
\includegraphics[width=7cm]{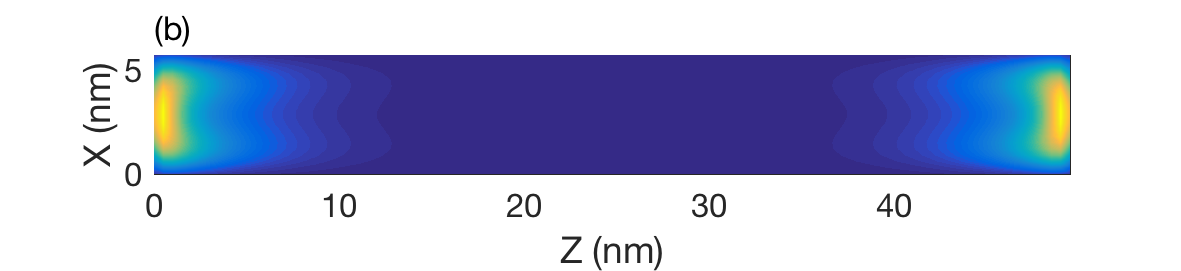}
\caption{(a) Energy structure of a  $6\times 50$ nm$^2$ ribbon. Each mode is doubly degenerate.(b) Cross-sectional LDOS distribution in the Dirac point ($E_D=0.295$ eV) in the ribbon.}
\label{rod1D}
\end{center}
\end{figure}

Fig.~\ref{rod1D}(a) shows the energy structure of a ribbon with the critical thickness $L_x=6$~nm  corresponding to the deepest gorge. 1D edge state with Dirac-like spectrum is clearly seen.  Fig.~\ref{rod1D}(b) shows the 
spacial distribution of LDOS across the sample cross-section near the ribbon side face ($x$-$y$ plane). Here the energy corresponds to the Dirac point. 
Thus, 1D states with Dirac energy spectrum develop near the edges of a thin ribbon in a proper thickness region. So such a ribbon can be considered a 2D topological insulator. 

In a ribbon of finite width, hybridization of the 1D edge states  results in the energy gap shown in  Fig.~\ref{decay}. Very slow decay of the gap with increase of ribbon width corresponds to $\approx 5$ times larger decay length in comparison with the surface states. In practice, such a slow decay means the properties of the 2D topological insulator can be clearly observed only in relatively large flakes with sizes in the range of tens of nanometers.

\subsection{Surface step}
The most experimentally relevant object is a surface step. Two types of steps  are analyzed: low (step height $L_s\lesssim \lambda$) and high ones  ($L_s\gg \lambda$). 

\begin{figure}
\begin{center}
\includegraphics[width=7cm]{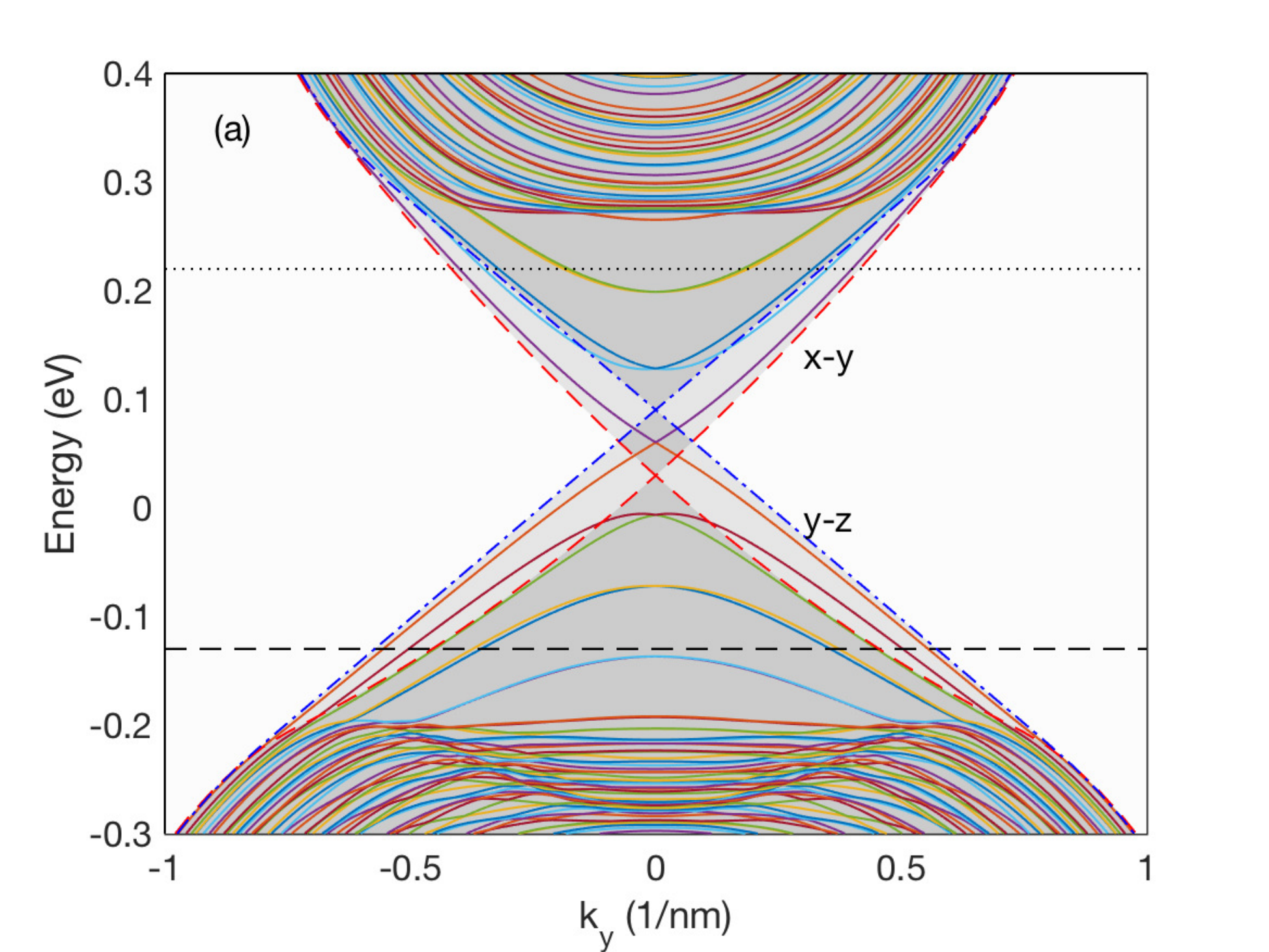}

\includegraphics[width=3.2cm]{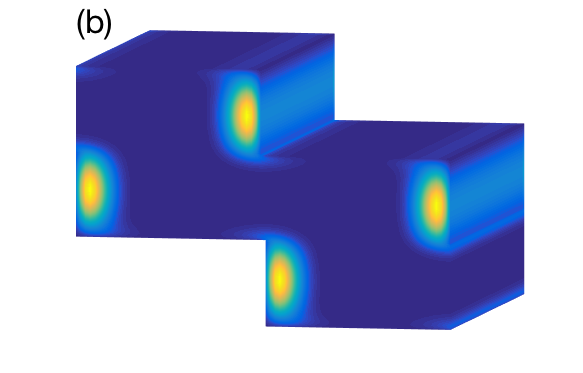}
\includegraphics[width=3.2cm]{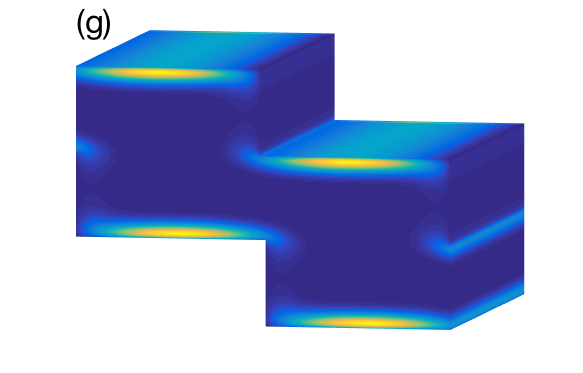}
\vskip -.2cm
\includegraphics[width=3.2cm]{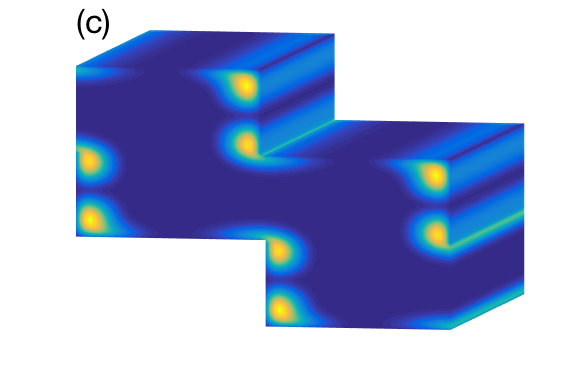}
\includegraphics[width=3.2cm]{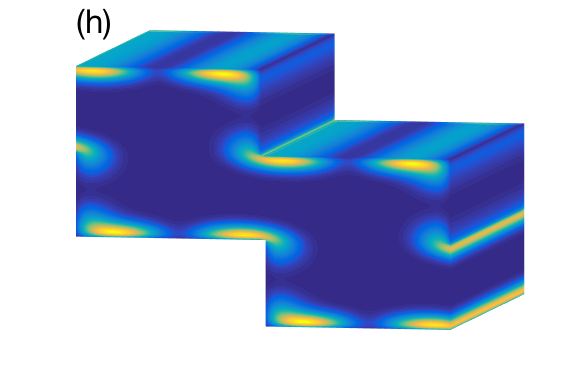}
\vskip -.2cm
\includegraphics[width=3.2cm]{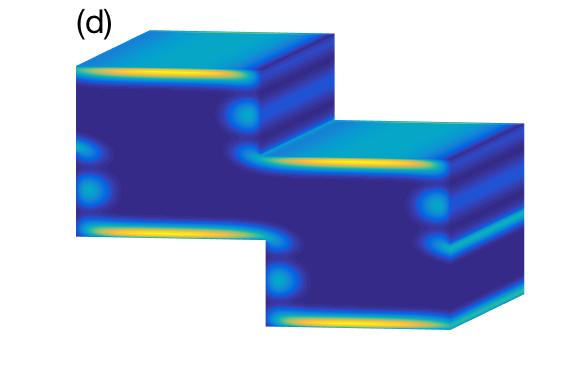}
\includegraphics[width=3.2cm]{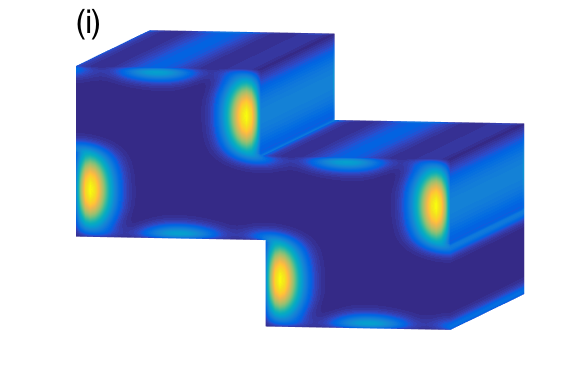}
\vskip -.2cm
\includegraphics[width=3.2cm]{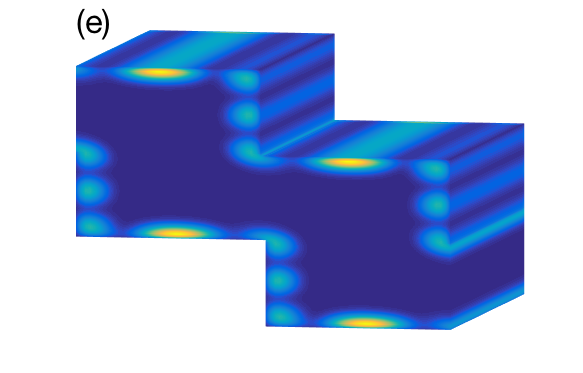}
\includegraphics[width=3.2cm]{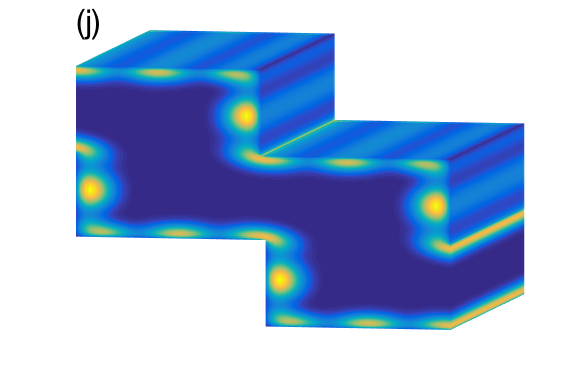}
\vskip -.2cm
\includegraphics[width=3.2cm]{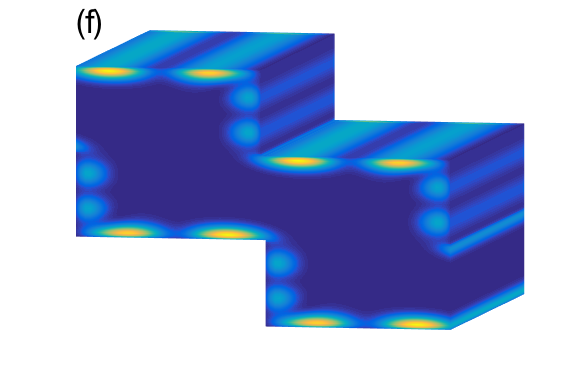}
\includegraphics[width=3.2cm]{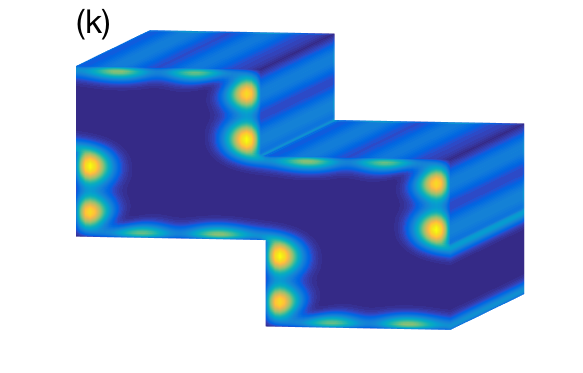}
\vskip -.2cm
\includegraphics[width=3.2cm]{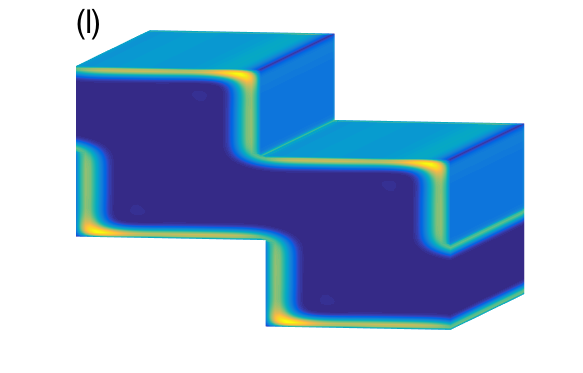}
\caption{a) Energy dispersion curves for stepped surface (step width 20 nm, step height 10 nm). Each mode is doubly degenerate. Dashed and dash-dotted lines correspond to $E(k_x=0,k_y)$ and $E(k_y,k_z=0)$ for $x$--$y$ and $y$--$z$ faces respectively. (b)--(l) Distribution of the partial LDOS  for electron states with $E=-0.13$ eV (dashed line in panel (a)) and $k_y=0.555$~nm$^{-1}$~(b), $0.503$~nm$^{-1}$~(c), $0.449$~nm$^{-1}$~(d), $0.372$~nm$^{-1}$~(e), $0.355$~(f)~nm$^{-1}$ and $E=0.22$~eV (dotted line in panel (a)) and $k_y=0.401$~nm$^{-1}$~(g), $0.351$~nm$^{-1}$~(h), $0.321$~nm$^{-1}$~(i), $0.188$~nm$^{-1}$~(j), $0.178$~nm$^{-1}$~(k), lowest panel, Dirac point).}
\label{lstepfig3}
\end{center}
\end{figure}

Fig. \ref{lstepfig3}(a) shows energy dispersion curves of  a slab hosting high surface steps. The energy spectrum is more complex in comparison with geometries described above. It demonstrates the same  Dirac cone-like mode as  a flat surface (Fig.~\ref{widefig3}(a)) and quantized states as in a rod (Fig.~\ref{rodfig3}). The survival of the Dirac cone is a consequence of the absence of the Berry phase contribution in this case, in contrast to  the case of the rod. 
The quantized states are now split in two. The splitting is caused by the difference of surface states dispersion on the top and the side surface of the step. 

Fig.~\ref{lstepfig3}(b)--(l) show partial LDOS obtained for different components of the spectrum below the Dirac point (left set of plots), above the Dirac point (right set of plots) and in the Dirac point (lowest panel). The features of LDOS resemble the ones described above for the rod. Namely, there are modes belonging only to a certain face ((b), (c), (g), (h)) as well as modes distributed over all faces. Again as in the case of the rectangular rod the face-specific modes can be identified as the ones lying inside the surface states cone of one face and outside of the surface states cone of the other face (light gray area in Fig.~\ref{lstepfig3}(a)). 
We also see that there is a mode with the Dirac spectrum, which is non-uniformly spread over the surface. 

\begin{figure}
\begin{center}
\includegraphics[width=7cm]{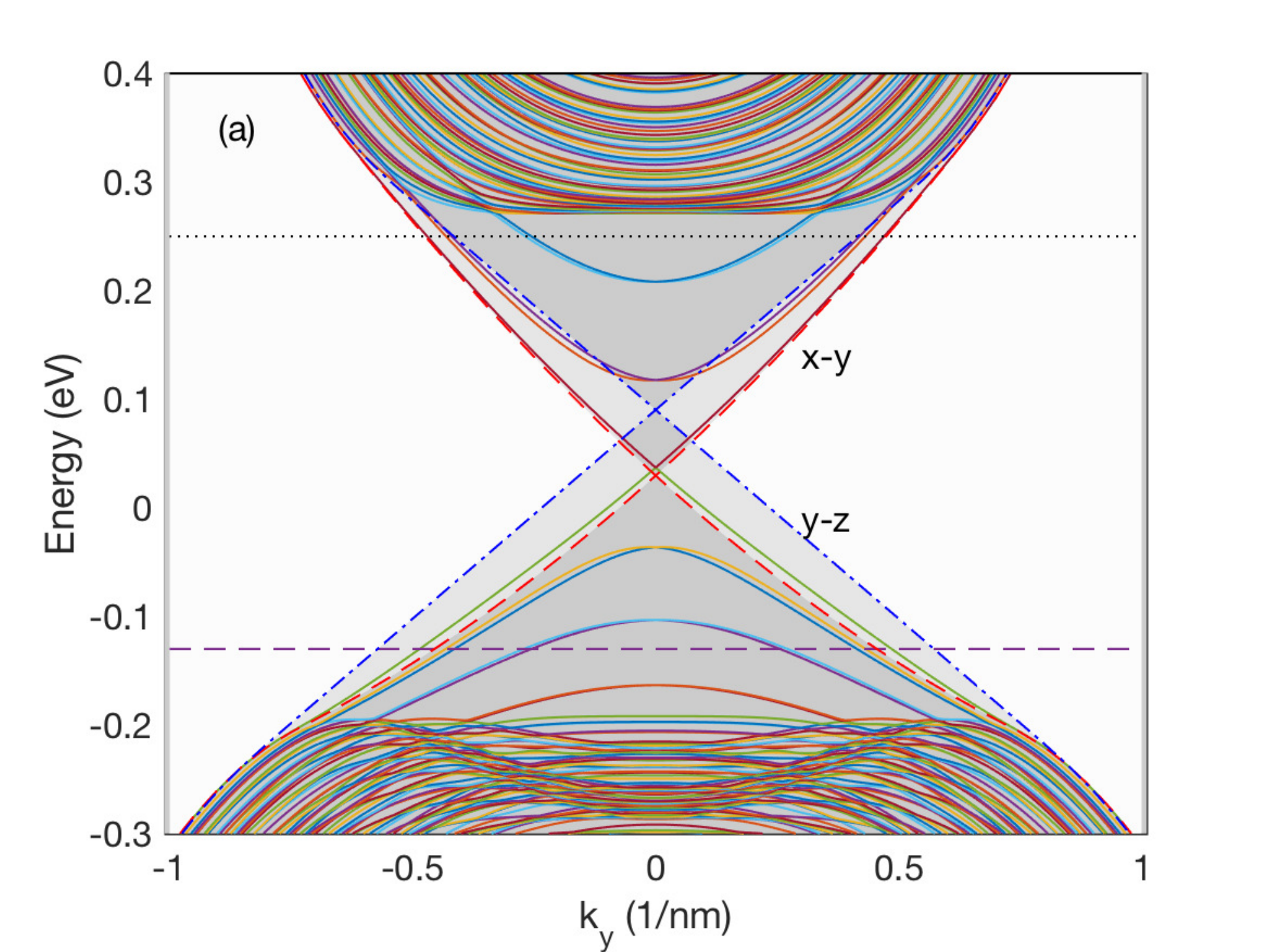}
\includegraphics[width=4cm]{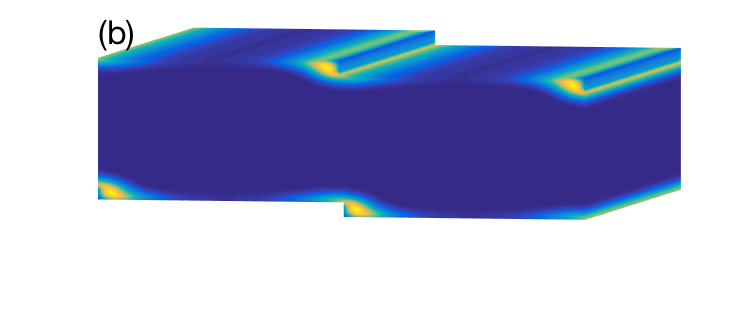}
\includegraphics[width=4cm]{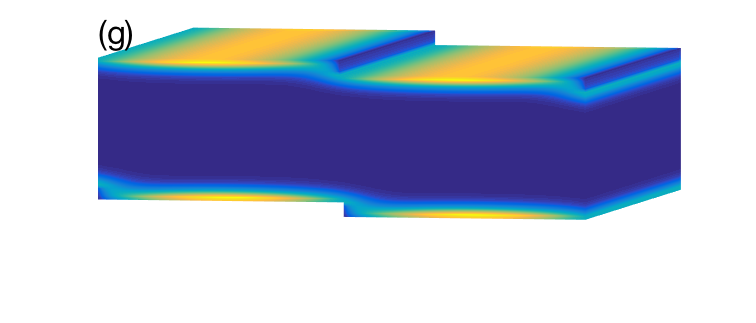}
\vskip -2mm
\includegraphics[width=4cm]{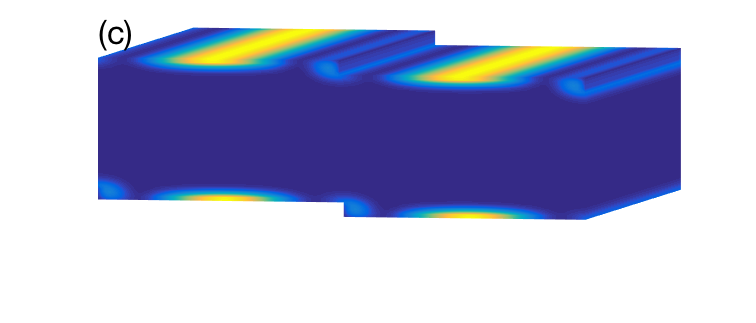}
\includegraphics[width=4cm]{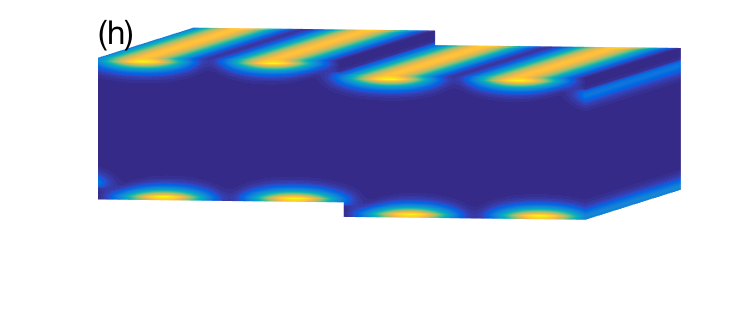}
\vskip -2mm
\includegraphics[width=4cm]{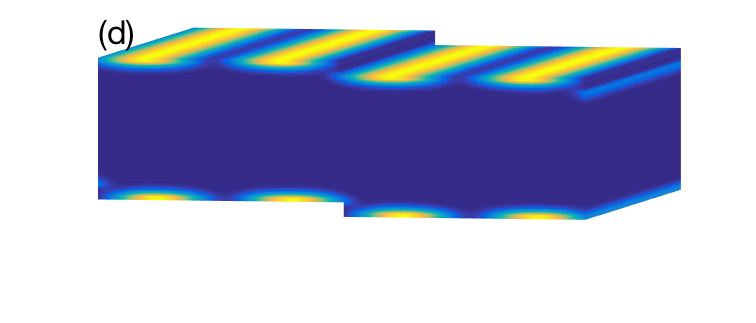}
\includegraphics[width=4cm]{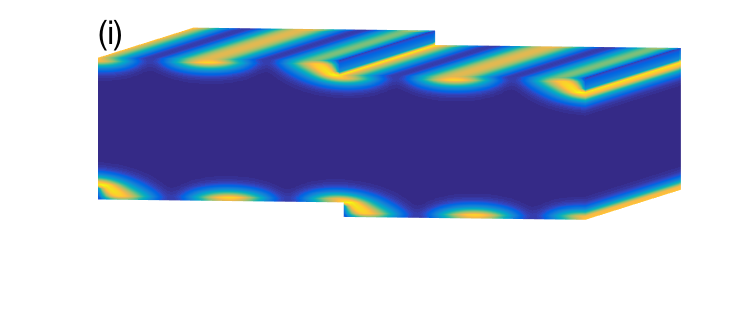}
\vskip -2mm
\includegraphics[width=4cm]{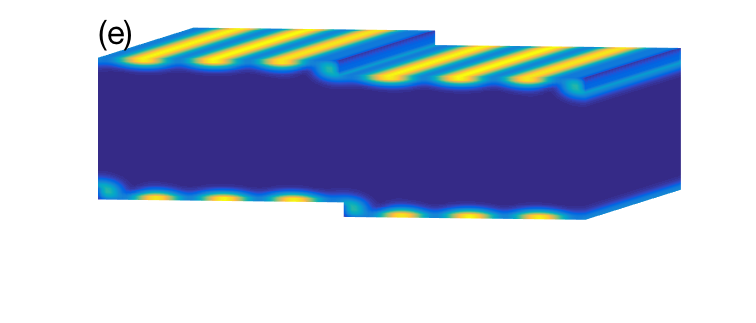}
\includegraphics[width=4cm]{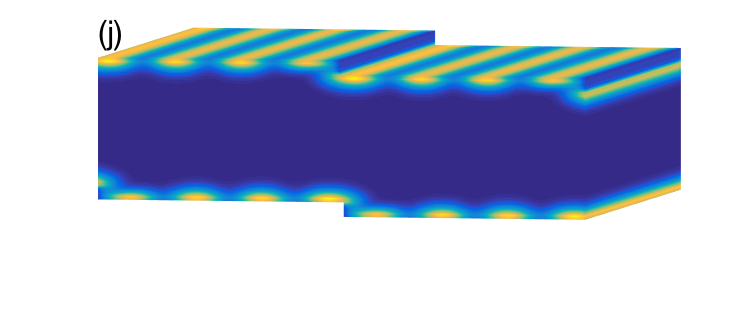}
\vskip -2mm
\includegraphics[width=4cm]{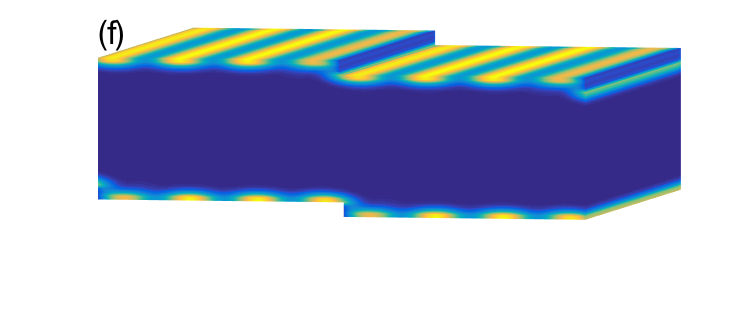}
\includegraphics[width=4cm]{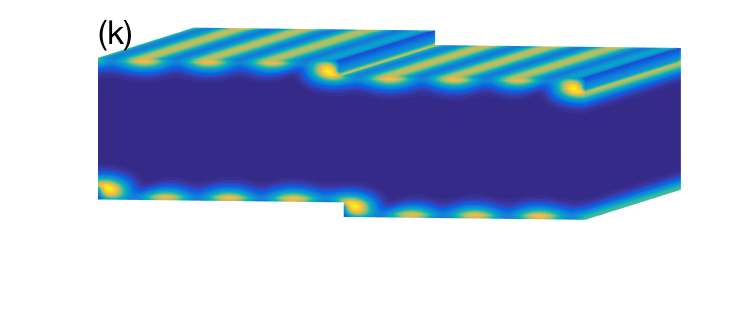}
\caption{(a) Energy dispersion curves for stepped surface with low steps (step width 30 nm, step height 2 nm) Each mode is doubly degenerate. 
Dashed and dash-dotted lines correspond to $E(k_x=0,k_y)$ and $E(k_y,k_z=0)$ for $x$--$y$ and $y$--$z$ faces respectively.(b)--(k) Cross-sectional partial LDOS distribution for different modes at electron energies $E=-0.13$ eV (dashed line) in a  slab with stepped surface. $k=0.485$~nm$^{-1}$~(b), $0.432$~nm$^{-1}$~(c), $0.413$~nm$^{-1}$~(d), $0.270$~nm$^{-1}$~(e), $0.255$~nm$^{-1}$~(f) and 
 for $E =  0.25$ eV (dotted line) and $k = 0.471$~nm$^{-1}$~(g), $0.432$~nm$^{-1}$~(h), $0.421$~nm$^{-1}$~(i), $0.269$~nm$^{-1}$~(j), $0.256$~nm$^{-1}$~(k)
}
\label{sstepfig}
\end{center}
\end{figure}

Very similar behavior is observed in a slab with low steps (Fig. \ref{sstepfig}). The Dirac mode is also present in the spectrum. However, no quantization along the z direction is observed. Splitting of surface states is smaller now. 

\begin{figure}
\begin{center}
\includegraphics[width=7cm]{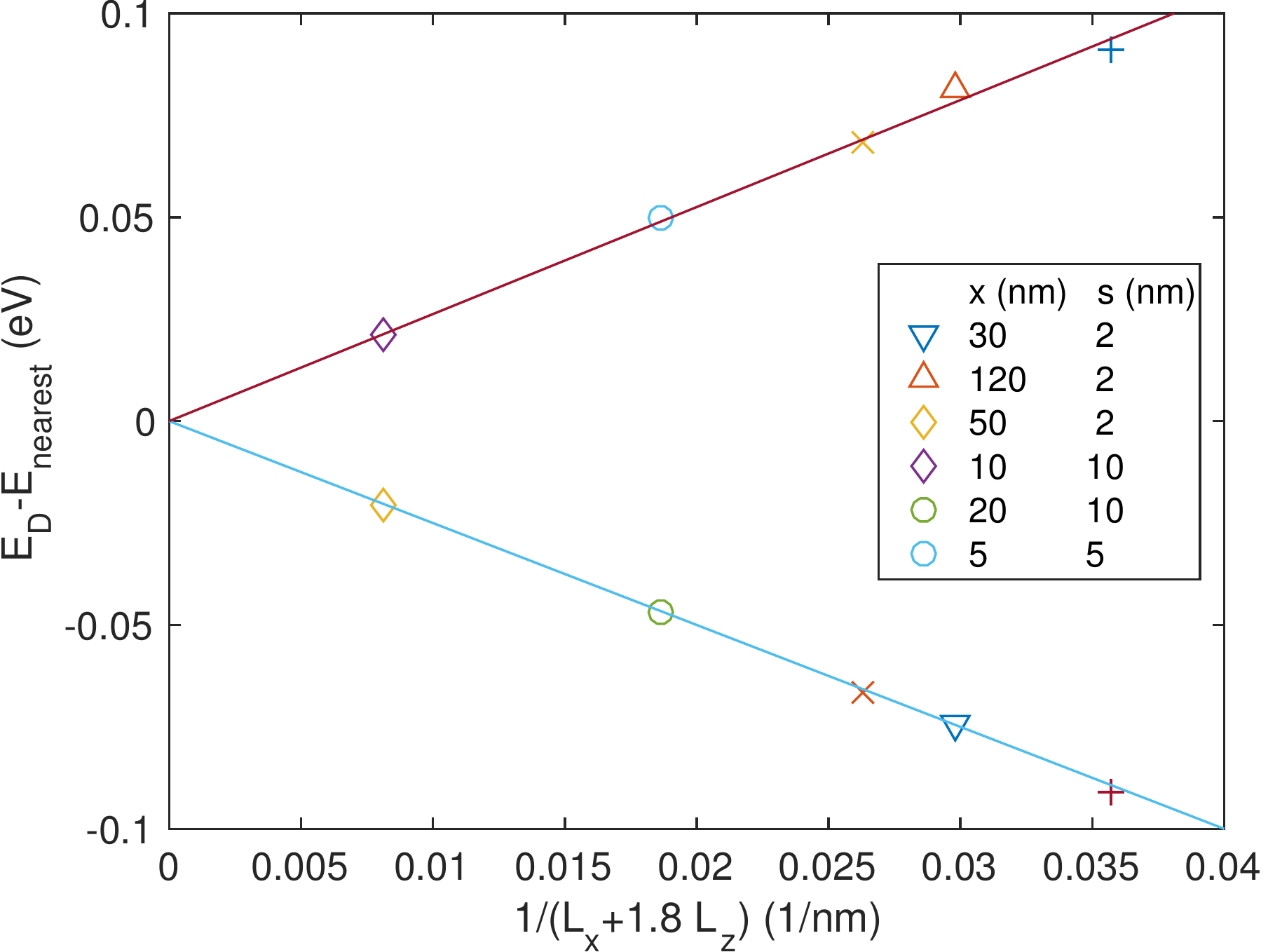}
\caption{Energy difference between the surface states quantization levels at $k_y=0$ as a function of inverse value of the effective length $L_{eff}=L_x+1.8L_s$ for stepped surface with periodical boundary conditions.}
\label{period}
\end{center}
\end{figure}

Fig. \ref{sstepfig}(b)-(k) show partial LDOS for energies below (left set) and above (right set) the Dirac point. The overall behavior resembles the one found in a slab with high steps. The difference is the absence of clear localization of states to a certain face. 
Fig.~\ref{period} shows the energy difference between the quantization levels at $k_y=0$ as a function of inverse effective surface length $L_{eff}=L_x + 1.8 L_s$ where $L_s$ is the step height. We see that resulting dependence is linear despite very different $L_x/L_s$ proportion, in agreement with equation~\ref{eq:delta}.

\begin{figure}
\begin{center}
\includegraphics[width=7cm]{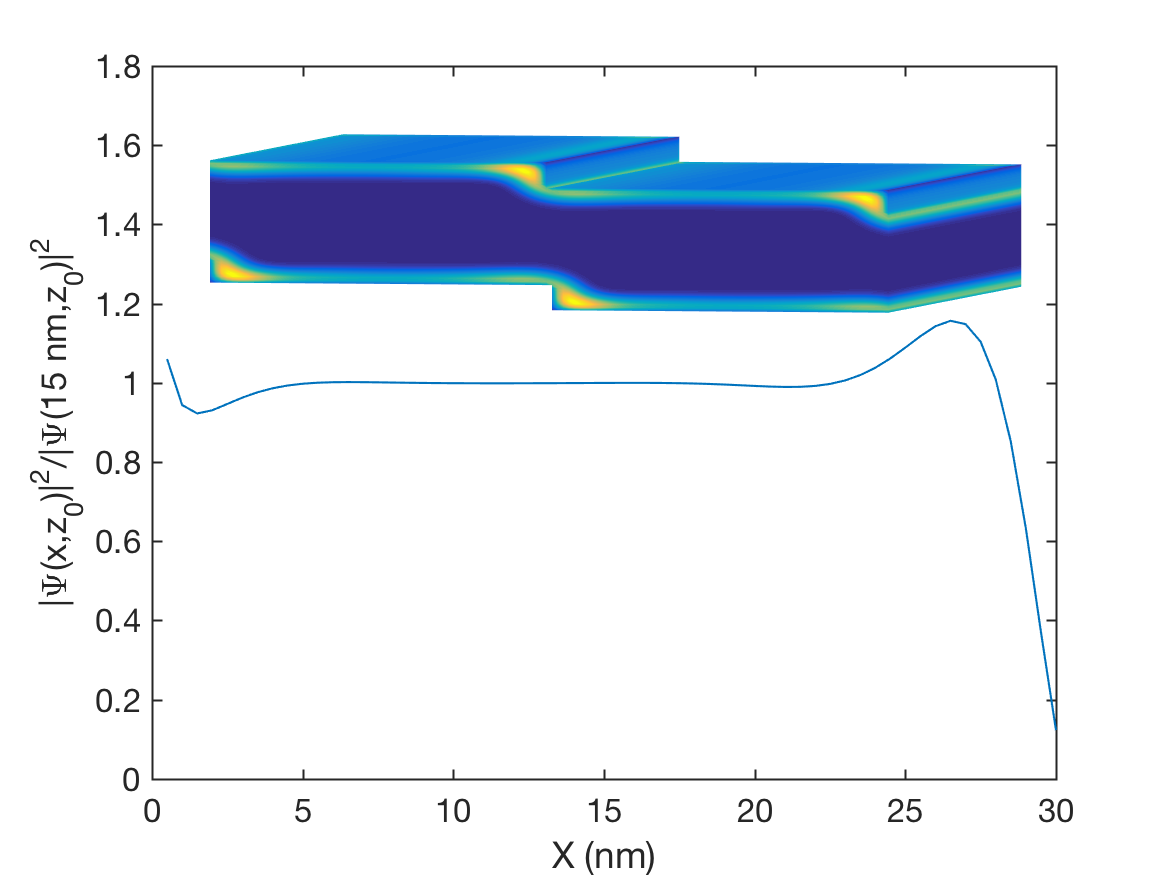}
\caption{ LDOS in $z_0=0.25$~nm surface layer as a function of the distance to the step edge averaged over a 5 meV energy region at the Dirac point ($E_D=36.4$~meV). To enhance the spatial resolution, the calculations were done for a stepped slab with 2 nm steps, step separation 30 nm and thickness 10 nm. The energy spectrum of this slab (especially for the surface states) is almost identical to one shown in Fig.~\protect\ref{sstepfig}(a).
Inset: cross-sectional LDOS distribution at the Dirac point for the same step. }
\label{sstep_ampl}
\end{center}
\end{figure}

One of the central questions of the present analysis is the prediction of the model for scanning tunneling spectroscopy near a step edge. Fig.~\ref{sstep_ampl} shows LDOS in a thin surface layer as a function of the distance from the surface step taken in the energy intervals around the Dirac point. We see that despite  the presence of modes highly localized near the step edges at certain energies (see Fig.~\ref{sstepfig}(b)), LDOS of the surface layer exhibits only a slight variation ($\lesssim 20$\%) near the step, in  agreement with the estimates described in \cite{Alpichshev}. This result supports also a conclusion of Ref.~\cite{FedotovPRB} that the main contribution to the increase of LDOS near the surface step in Bi$_2$Se$_3$ found in tunneling experiments comes from the effect of the chemical potential shift \cite{FedotovJETPL} rather than from formation of one-dimensional states near the step edge.

 \section{Summary}
We see that the effective continuous model  \cite{Zhang2009} reproduces a number of well-known features of topological insulators,  namely the presence of the surface states with the same depth distribution, as {\em ab initio} calculations \cite{Eremeev},
different dispersion laws and Dirac point positions on  different crystal faces also known from {\em ab initio} calculations \cite{Moon},  oscillatory character of the band structure {\em vs.} slab thickness \cite{oscillations,oscillations1}. Thus this model can serve as an effective  tool for  analysis of topological insulator nanostructures. Its convenience is due to a small number of parameters and the possibility to model large nanostructures.

The model predicts a 2D topological insulator state in slabs oriented not only along the $x$ and $y$ but also along the $z$ axis.
No edge states localized near rod's edges or at the step edge appear in the framework of this model.
We find instead modes localized on different faces of the rod running along the $y$ direction. Such states appear also in tight binding calculations \cite{Tightbinding}. We argue that the modes are massive and their origin  is due to the difference  in the Dirac point energy of adjacent faces. No such states are found in the    rod running along the $z$ direction, as the surface states of its faces have the same  Dirac point position.
 
For the large steps running along the $y$ direction we find modes localized at the top or side surface of the step similar to the case of the rod.
The wave functions on the small steps behave in a different manner. The tendency of the modes to localize on one or the other step is counteracted by the tendency of the surface states to spread out, which is made possible by the fact that the step height is smaller than the penetration depth of the surface states.
At the same time, the small increase of the LDOS near the step  \cite{Alpichshev} is reproduced by this model. A decrease of LDOS is observed near the concave part of the steps.
In a real situation the difference in Dirac point position on different surfaces would lead to a potential difference and hence to redistribution of the electron density \cite{Silvestrov}, including possibly formation of bound states \cite{FZZ2018}.

\ack
Financial support from RScF  (project \# 16-12-10335) is acknowledged.

\end{document}